\documentclass[aps,amsfonts,epsf]{revtex4}
\usepackage{graphics}
\usepackage{graphicx}
\usepackage{epsfig}

\def\beq{\begin{equation}}
\def\eeq{\end{equation}}
\def\bea{\begin{eqnarray}}
\def\eea{\end{eqnarray}}

\def\reff@jnl#1{{\rm#1\/}}
\def\prd{\reff@jnl{Phys. Rev. D }}        
\def\cqg{\reff@jnl{Class. Quantum Grav. }} 
\def\ekp{Porter~E~K }
\def\njc{Cornish~N~J }
\def\jrg{Gair~J~R }
\begin{document}
\input epsf.tex

\title{A Hamiltonian Monte Carlo method for Bayesian Inference of Supermassive Black Hole Binaries.}
\author{Edward K. Porter \& J\'er\^ome Carr\'e } 
\affiliation{Francois Arago Center, APC, UMR 7164, Universit\'e Paris 7 Denis Diderot,\\
                10, rue Alice Domon et L\'eonie Duquet, 75205 Paris Cedex 13, France}
\vspace{1cm}

\vspace{1cm}
\begin{abstract}
We investigate the use of a Hamiltonian Monte Carlo to map out the posterior density function for supermassive black hole binaries.  While
previous Markov Chain Monte Carlo (MCMC) methods, such as Metropolis-Hastings MCMC, have been successfully employed for a number
of different gravitational wave sources, these methods are essentially random walk algorithms.  The Hamiltonian Monte Carlo treats the inverse
likelihood surface as a ``gravitational potential" and by introducing canonical positions and momenta, dynamically evolves the Markov chain by solving
Hamilton's equations of motion.  We present an implementation of the Hamiltonian Markov Chain that is faster, and more efficient by a factor of approximately
the dimension of the parameter space, than the standard MCMC.
\\

\end{abstract}

\maketitle

\section{Introduction}
The inspiral and merger of supermassive black hole binaries (SMBHBs) are expected to be some of the most violent, and in terms of gravitational waves (GW),
brightest events in the Universe.  A space-based observatory, such as eLISA, has the potential of detecting between 10 -100 events per year, with a signal-to-noise
ratio (SNR) $>$ 10, out to a redshift of $z\sim 20$~\cite{eLISA}.   While we do not expect there to be any significant confusion between various signals, the waveform
parameters themselves can be highly correlated.  In the early days of GW astronomy, the Fisher Information Matrix (FIM) was the standard tool for the estimation of parameter
errors.  In the high SNR limit, this seemed a valid approximation and one could easily and quickly invert the FIM to attain the variance-covariance matrix.  However, as the FIM is a local quadratic approximation to the likelihood, it does not provide information on other modes of the solution.  It was also demonstrated  that, for example, due to symmetries in the waveform, bi-modal solutions exist due to reflections in sky position, and due to the wave-like nature of the signals, whole island chains of solutions resulting from the waveforms going in and out of phase with the true signal~\cite{B1}.  Furthermore,
as the FIM is almost always nearly singular, it has difficulties dealing with numerical instabilities, especially if these arise from strong correlations between parameters.  Therefore, it is not always clear if the result we get from the numerical inversion is something that can be trusted.

More recently, Bayesian analysis has become a very useful tool in the field of GW astronomy.  In recent years, for space-based observatories,
this method has been applied to SMBHBs, galactic binaries, cosmic (super)strings, EMRIs and stochastic cosmological backgrounds~\cite{B1,B2,B3,B4,B5,B6,B7,B8,B9,B10,B11,B12,B13,B14,B15,B16,B17,B18,B19,B20,B21,B22,B23}.  Many of these studies have demonstrated the power
of algorithms such as Markov Chain Monte Carlo (MCMC) in mapping the posterior density functions (pdf) for waveform parameters.  While the MCMC takes much
longer to run than the calculation of a FIM, if done properly, the estimations coming from the chains are more trustworthy.  The downside to MCMC algorithms is
their implementation.  The standard algorithms such as Gibb's sampling~\cite{gibbs} and Metropolis-Hastings~\cite{MH1,MH2} MCMC algorithms are both examples of
what are called random walk MCMCs.   These chains move around the equilibrium distribution by taking small steps in the parameter space.  By randomly proposing jumps in the parameter space, the algorithm then invokes a predetermined decision process to decide whether of not to accept the new point.  While MCMCs are guaranteed to converge 
to the distribution we are looking to explore, there is no guarantee of how long it will take to do so.  Therefore,  not only can the choice of proposal distribution can 
greatly accelerate the convergence of the chain, but so too can the size of the steps taken.  The required axiom of ``properly sized steps in the right direction" effectively
means that the MCMC has to be tuned to the problem at hand.  Furthermore, we expect a high level of autocorrelation between samples in the chains.  This means
that we always have to be sure that the chains are run long enough that the autocorrelations have been sufficiently reduced.

An alternative to the random walk MCMC was proposed by Duan \emph{et al}~\cite{HMC} where, by introducing a set of canonical momenta and state space 
variables, one evolves the Markov chain by simulating Hamiltonian dynamics.  Rather than a random sampling of the parameter space, here one inverts the
equilibrim distribution and uses gradient information, combined with the canonical momenta, to evolve trajectories in phase space.  As the Hamiltonian
is not completely conserved, a Metropolis-Hastings step is introduced at the end of each trajectory to make the comparison with the starting point.  By following
trajectories in phase space, we not only explore distant points in the parameter space quicker, but we reduce the random nature of the chain which improves the
chain mixing, resulting in a higher acceptence rate compared to a standard random walk MCMC.  It has been shown that a Hamiltonian Markov Chain (HMC) of 
a certain length can provide the same efficiency as a random walk MCMC which is $D$ times longer, where $D$ is the dimensionality of the parameter space~\cite{Hajian}.  With these apparant benefits, one might ask why the HMC is not the standard MCMC of choice?  The answer for this is simple : as with any such
algorithm, there are a number of internal parameters that need to be tuned, such as the time-step in the Hamiltonian dynamics and the the length of the 
trajectories.  And then finally, there is the question of the gradients needed to evolve Hamilton's equations.  For many astrophysical applications, the distribution
to be mapped is the likelihood distribution.  In many problems, the calculation of the likelihood is also computationally expensive.  To implement the HMC one needs 
to calculate the gradient of the log likelihood at each step of the trajectory.  Unless one has a very simple and/or possibly analytical likelihood to calculate, these
gradient calculations can very easily create a bottleneck for the implementation of this algorithm.

For GW astronomy, as there is no closed analytical form to the calculation of the log likelihood, our only option is a numerical calculation.  This usually involves
the generation of waveforms, which can be time consuming.  In this article we first demonstrate a method for tuning the step size in the discrete Hamiltonian 
equations, as well as the trajectory length.   We then demonstrate that by using information from a relatively small number of  initial trajectories, we can use an
 analytic cubic function to individually fit the log likelihood gradients in the parameter coordinates, which dramatically improves the run-time of the code.

\subsection{Outline of the paper}
The paper is structured as follows.  In Section~(\ref{sec:gw}) we define the gravitational waveform for non-spinning black hole binaries, the response of a detector to
such waveforms in the low frequency approximation and some of the GW tools that are needed for the analysis.  In Section~(\ref{sec:mcmc}) we define MCMC algorithms
such as Gibb's sampling and Metropolis-Hastings MCMC.  We finish with a definition of the Hamiltonian Markov Chain algorithm.  In Section~(\ref{sec:hmc}) we look
at the methods used to improve the efficiency of the HMC algorithm.  Section~(\ref{sec:mvh}) contains an initial comparison between the MCMC and HMC
algorithms and study a method for reducing the bottleneck in calculating the gradients of the log likelihood.  This is followed in Section~(\ref{sec:results}) where
we conduct a full comparison between the two algorithms.

\section{The Gravitational Waveform and Tools for Analysis}\label{sec:gw}
The strain of the gravitational wave (GW) in each channel of a LISA-type detector~\cite{lisa}, with both polarizations is given by
\begin{equation}
h(t) = h_{+}(\xi(t))F^{+}+h_{\times}(\xi(t))F^{\times}, 
\label{eq:hoft}
\end{equation}
where the phase shifted time parameter is 
\begin{equation}
\xi(t) = t - R_{\oplus}\sin\theta\cos\left(\alpha(t) - \phi\right).
\end{equation}
Here, $R_{\oplus} = 1 AU \approx$ 500 secs is the radial distance to the detector guiding center, $\left(\theta,\phi\right)$ are the position angles of the source in the sky, $\alpha(t)=2\pi f_{m}t + \kappa$, $f_{m}=1/year$ is the observatory modulation frequency and $\kappa$ gives the initial ecliptic longitude of the guiding center.  Using a restricted post-Newtonian approximation, the GW polarizations are given by
\begin{eqnarray}
h_{+} &=& \frac{2Gm\eta}{c^{2}D_{L}}x\,\cos\left(2\Phi_{orb}(\xi)\right),\nonumber \\ \nonumber \\
h_{\times} &=& -\frac{4Gm\eta}{c^{2}D_{L}}x\,\sin\left(2\Phi_{orb}(\xi)\right),
\label{eqn:strain}
\end{eqnarray}
where the total mass of the binary is $m=m_{1}+m_{2}$, $\eta = m_{1}m_{2}/m^{2}$ is the reduced mass ratio and $D_{L}$ denotes the luminosity distance.  The invariant PN velocity parameter is defined by, $x = \left(Gm\omega / c^{3}\right)^{2/3}$, where $\omega=d\Phi_{0rb}/dt$ is the 3 PN order orbital frequency for a circular orbit and $\Phi_{orb}=\varphi_{c}^{orb}-\phi_{orb}(t)$ is the orbital phase which is defined as
\bea
\label{eq:SMBH_phase}
\phi_{3PN}(t) & = & \frac{\varphi_c}{2} - \frac{1}{\eta} \left[
		 \Theta^{5/8} 
		 + \left(\frac{3715}{8064}+\frac{55}{96}\right)\Theta^{3/8}
		  - \frac{3\pi}{4}\Theta^{1/4}+\left(\frac{9275495}{14450688}+\frac{284875}{258048}\eta+\frac{1855}{2048}\eta^2\right)\Theta^{1/8} \right. \nonumber \\
		  & + & 
		    \left(\frac{38645}{21504}-\frac{65}{256}\eta\right)\ln\left(\frac{\Theta}{\Theta_{lso}}\right)\pi + \left\{ \frac{831032450749357}{57682522275840}-\frac{53}{40}\pi^2 + \left(-\frac{126510089885}{4161798144}+\frac{2255}{2048}\pi^2\right)\eta-\frac{107}{56}\gamma \right. \nonumber \\
		  & + & \left. \frac{154565}{1835008}\eta^2-\frac{1179625}{1769472}\eta^3-\frac{107}{56}\ln(2\Theta)\right\} \Theta^{-1/8} +  \left. \left(\frac{188516689}{173408256}+\frac{488825}{516096}\eta-\frac{141769}{516096}\eta^2\right)\pi\Theta^{-1/4} \right],
\label{eq:phase_SMBH}
\eea
and the quantity $\Theta$ is related to the time to coalescence of the wave, $t_{c}$, by
\begin{equation}
\Theta(t) = \frac{c^{3}\eta}{5Gm}\left(t_{c}-t\right).
\end{equation}
$\varphi_{c}^{orb}$ is the constant orbital phase of the wave at coalescence and $\gamma$ is Euler's constant.  All GW phases are then twice the orbital value.  To account for cases where the above phase equation diverges before we reach the last stable orbit, we truncate our waveform generation at $R=7M$.

Using the concurrent WMAP values of $(\Omega_{R}, \Omega_{M}, \Omega_{\Lambda}) = (4.9\times10^{-5}, 0.27, 0.73)$ and a Hubble's constant of $H_{0}$=71 km/s/Mpc~\cite{Verdi}, the relation between redshift, $z$, and luminosity distance, $D_{L}$, is given by
\begin{equation}
D_{L} = \frac{c(1+z)}{H_{0}}\int_{0}^{z}dz'\left[\Omega_{R}\left(1+z'\right)^{4}+\Omega_{M}\left(1+z'\right)^{3} + \Omega_{\Lambda}\right]^{-1/2}  .
\end{equation}

The functions $F^{+,\times}$ in Equation~(\ref{eq:hoft}) are the beam pattern functions of the detector given in the low frequency approximation by 
\beq
F^{+}(t;\psi, \theta, \phi, \lambda) = \frac{1}{2}\left[\cos(2\psi)D^{+}(t;\theta, \phi, \lambda) - \sin(2\psi)D^{\times}(t;\theta, \phi, \lambda)\right],
\eeq
\beq
F^{\times}(t;\psi, \theta, \phi, \lambda) = \frac{1}{2}\left[\sin(2\psi)D^{+}(t;\theta, \phi, \lambda) + \cos(2\psi)D^{\times}(t;\theta, \phi, \lambda)\right],
\eeq
where $\psi$ is the polarization angle of the wave and $\lambda = 0$ or $\pi/4$ defines the two-arm combination of a LISA-type observatory from which the strain is coming.  The detector pattern functions are given by~\cite{CR}
\bea
D^{+}(t) &=& \frac{\sqrt{3}}{64}\left[\frac{}{}-36\sin^{2}(\theta)\sin(2\alpha(t)-2\lambda)+(3+\cos(2\theta)) \left(\frac{}{}\cos(2\phi)\left\{\frac{}{}9\sin(2\lambda)-\sin(4\alpha(t)-2\lambda)\right\} \frac{}{}
\right.\right. \\ \nonumber
&+&\left.\left.\sin(2\phi)\left\{\frac{}{}\cos\left(4\alpha(t)-2\lambda\right)-9\cos(2\lambda) \right\}\frac{}{}\right)-4\sqrt{3}\sin(2\theta)\left(\frac{}{}\sin(3\alpha(t)-2\lambda-\phi)\right.\right.\\
&-&\left.\left.3\sin(\alpha(t)-2\lambda+\phi)\frac{}{}\right)\frac{}{}\right] \nonumber,
\eea
\bea
D^{\times}(t) = \frac{1}{16}\left[\frac{}{}\sqrt{3}\cos(\theta)\left(\frac{}{}9\cos(2\lambda-2\phi)-\cos(4\alpha(t)-2\lambda-2\phi) \right) \right. \\ \nonumber \left. -6\sin(\theta)\left(\frac{}{} \cos(3\alpha(t)-2\lambda-\phi)+3\cos(\alpha(t)-2\lambda+\phi) \right) \right].
\eea
\noindent The low frequency approximation is an extremely good fit to the full detector response at frequencies of $\lesssim$ 3 mHz for a detector with an armlength of $L=5\times10^9$ m~\cite{RPC}.  In this case the two channel formalism originally derived by Cutler~\cite{cc} corresponds to the construction of optimal orthogonal time delay interferometry (TDI) variables $\{A,E\}$ using the unequal-arm Michelson TDI variables $\{X,Y,Z\}$ according to 
\begin{equation}
A = X\,\,\,\,\,\,\, ,\,\,\,\,\,\, E = (X+2Y) / \sqrt{3}.
\end{equation}

We now present some of the tools needed for the future analysis of the algorithms.  The first thing we define is the inner product between two 
real functions $h(t)$ and $s(t)$ as
\begin{equation}\label{eqn:scalarprod}
\left<h\left|s\right.\right> =2\int_{0}^{\infty}\frac{df}{S_{n}(f)}\,\left[ \tilde{h}(f)\tilde{s}^{*}(f) +  \tilde{h}^{*}(f)\tilde{s}(f) \right],
\label{eq:scalarprod}
\end{equation}
where a tilde denotes a Fourier transform and an asterisk denotes a complex conjugate.  The denominator $S_{n}(f)=S_{n}^{instr}(f)+S_{n}^{conf}(f)$ is the one-sided noise spectral density of the detector, which is a combination of instrumental and galactic confusion noise.  For the instrumental noise we use the expression given by
\begin{equation}
S_{n}^{instr}(f)=\frac{1}{4L^{2}}\left[ 2 S_{n}^{pos}(f)\left(2+\cos^{2}\left(\frac{f}{f_{*}}\right)\right)+8 S_{n}^{acc}(f)\left(1+\cos^{2}\left(\frac{f}{f_{*}}\right) \right)\left(\frac{1}{(2\pi f)^{4}}+\frac{\left(2\pi 10^{-4}\right)^2}{(2\pi f)^{6}}\right)\right] , 
\end{equation}
where $L=5\times10^{6}$ km is the arm-length for a LISA-type detector,  $S_{n}^{pos}(f) = 4\times10^{-22}\,m^{2}/Hz$ and $S_{n}^{acc}(f) = 9\times10^{-30}\,m^{2}/s^{4}/Hz$ are the position and acceleration noise respectively.  The quantity $f_{*}=1/(2\pi L)$ is the mean transfer frequency for the detector arm.  Notice that the final term in the expression has the effect of reddening the noise below $10^{-4}$ Hz to account for the fact that we may not be able to achieve the desired noise spectral density as we approach $10^{-5}$ Hz. For the galactic confusion we use the following confusion noise estimate derived from a Nelemans, Yungelson, Zwart (NYZ) galactic foreground model~\cite{nyp, trc}
\begin{equation}
S_{n}^{conf}(f) = \left\{ \begin{array}{ll} 10^{-44.62}f^{-2.3} & 10^{-4} < f\leq 10^{-3} \\ \\ 10^{-50.92}f^{-4.4} & 10^{-3} < f\leq 10^{-2.7}\\ \\ 10^{-62.8}f^{-8.8} &  10^{-2.7} < f\leq 10^{-2.4}\\ \\ 10^{-89.68}f^{-20} &  10^{-2.4} < f\leq 10^{-2}  \end{array}\right.,
\end{equation}
where the confusion noise has units of $Hz^{-1}$.  

We also define the signal to noise ratio (SNR) for a source in each individual detector as
\begin{equation}
\rho_{i} = \frac{\left<s_{i}\left|h_{i}\right.\right>}{\sqrt{\left<h_{i}\left|h_{i}\right.\right>}}.
\end{equation}
When we use both detectors the total SNR is given by
\begin{equation}
\rho = \sqrt{\rho_{I}^2 + \rho_{II}^2},
\end{equation}

Finally, for the particular problem of SMBHBs, it is convenient to work with the parameter set $x^{\mu} = \{\ln(M_{c}), \ln(\mu), \ln(t_{c}), \cos\theta, \phi, \ln(D_{L}), \cos\iota, \varphi_{c}, \psi\}$ where $M_{c}=m\eta^{3/5}$ is the chirp mass, $\mu =m\eta$ is the reduced mass and all other parameters have been previously defined. 

\section{The Markov Chain Monte Carlo}\label{sec:mcmc}
When it comes to GW parameter estimation, the first port of call for many people has been the Fisher Information Matrix (FIM).  However, in GW astronomy, as the FIMs are always
almost singular, there can be problems with the numerical inversion, leading to spurious predictions.  A further problem is that results from the FIM do not properly reflect situations 
where there is an extremely high correlation in the system parameters or multi-modal solutions.  A viable alternative to the FIM is the use of a Markov Chain Monte Carlo (MCMC) algorithm to map out the posterior density function $\pi(\vec{x})$ of the source and estimate the precision in the parameter estimation..  The simplest MCMC is based on Gibbs sampling~\cite{gibbs}.  This is a simple MCMC to implement where moves are always accepted.  The algorithm works on the premise that instead of drawing from a full multivariate distribution, we use draws from univariate conditional distributions.  This way the algorithm simulates $n$ random variables sequentially from $n$ conditional distributions.  Gibbs sampling works on the premise, if some joint distribution $P(x,y)$ exists, it is easier to consider the conditional probabilities $P(x|y)$ and $P(y|x)$ (where $P(A|B)$ denotes the conditional probaility of A given B) rather than the marginal distribution $P(x) = \int P(x,y)dy$.  While simple to implement, the Gibbs sampling suffers from an achingly slow convergence, mostly due to the tendency to go on large random walks in the parameter space due to the fact that all proposals are accepted.  

\subsection{Metropolis-Hastings MCMC}
A better algorithm is based on a Metropolis-Hastings MCMC~\cite{MH1,MH2}.  In general, the algorithm works as follows.  Given a signal $s(t)$, the likelihood that the true parameter values are given by a D-dimensional parameter vector $x^{\mu}$ is given by
\beq
{\mathcal L}(x^{\mu}) = C\exp\left[-\frac{1}{2}\left< s-h(x^{\mu}) \left| s-h(x^{\mu})\right.\right>  \right],
\label{eq:likelihood}
\eeq 
where $h(x^{\mu})$ is our waveform model and $C$ is a normalization constant.  For SMBHBs, in the high SNR limit, we assume the errors in the estimation of the system parameters is described by the multivariate Gaussian probability distribution 
\beq
p(\Delta x^{\mu}) = \sqrt{\frac{|\Gamma|}{2\pi}}\exp\left[ -\frac{1}{2}\Gamma_{\mu\nu}\Delta x^{\mu}\Delta x^{\nu}\right],
\eeq 
where we define the FIM
\beq
\Gamma_{\mu\nu} = \left<\frac{\partial h}{\partial x^{\mu}}\left|\frac{\partial h}{\partial x^{\nu}}\right. \right> = -E\left[ \frac{\partial^2 \ln {\mathcal L}}{\partial x^{\mu} \partial x^{\nu}}\right],
\label{eq:FIM}
\eeq
and $|\Gamma| = det(\Gamma_{\mu\nu})$.  We can see here that there is a relation between the FIM and the negative expectation value of the Hessian matrix of the log likelihood.
In order to explore the posterior density $\pi(x^{\mu})$ the method works as follows : starting with the signal $s(t) = h(t) + n(t)$, where $n(t)$ is the noise in the detector, and some initial model template $h(t;x^{\mu})$ constructed by choosing a random starting point in the parameter space $x^{\mu}$, we then draw from a proposal distribution and propose a jump to another point in the space $y^{\mu}$, where $y^{\mu}=x^{\mu}+\Delta x^{\mu}$.  In order to compare both points, we evaluate the Metropolis-Hastings ratio
\begin{equation}
H = \frac{\Pi(y^{\mu})p(s|y^{\mu})q(x^{\mu}|y^{\mu})}{\Pi(x^{\mu})p(s|x^{\mu})q(y^{\mu}|x^{\mu})}.
\end{equation}
Here $\Pi(x^{\mu})$ are the prior distributions of the parameters and $p(s|x^{\mu})$ is the likelihood defined by Equation~(\ref{eq:likelihood}).  The quantity $q(x^{\mu}|y^{\mu})$ is the proposal distribution used for jumping from $x^{\mu}$ to $y^{\mu}$.  In order to make jumps in the parameter space, the most efficient proposal distribution to use is a multivariate Gaussian distribution.  The multivariate jumps use a product of normal distributions in each eigendirection of $\Gamma_{\mu\nu}$, with the standard deviation in each eigendirection given by $\sigma_{\mu} = 1/\sqrt{DE_{\mu}}$,  where $D$ is the dimensionality of the search space and $E_{\mu}$ are the eigenvalues of the FIM.  Once a jump is proposed, it is then accepted with probability $\alpha = min(1,H)$, otherwise the chain stays at $x^{\mu}$.  

\subsection{Hamiltonian Monte Carlo}
The likelihood function defined in Equation~(\ref{eq:likelihood}) has the form of a canonical partition function where we can write ${\mathcal L}(x^{\mu})=C \exp(-\beta E(x^{\mu}))$, where $\beta$ is an inverse temperature that takes the value of $\beta=1/2$ in this case and $E(x^{\mu})$ is an energy state.  Pushing this analogy further, we can think of exploring the posterior density function as an inverse ``gravitational potential well".  By inverting the log likelihood, the Hamiltonian Monte Carlo (HMC) then uses the gradient information in the inverse log likelihood surface to reduce the random walk nature of the MCMC.

Following Duan \emph{et al}~\cite{HMC}, by defining the parameter values as state space variables, i.e.  $q^{\mu} = x^{\mu}$, and augmenting these variables by a set of canonical momenta $p^{\mu}$ that define a kinetic energy term $K(p^{\mu}) = M_{\mu\nu}^{-1}p^{\mu}p^{\nu} / 2$, we can write down the Hamiltonian
\beq
{\mathcal H}(q^{\mu},p^{\mu})=E(q^{\mu})+K(p^{\mu})= -\ln \mathcal{L}(q^{\mu}) + \frac{1}{2}  M_{\mu\nu}^{-1}p^{\mu}p^{\nu},
\eeq
where we assume $-\ln \mathcal{L}(q^{\mu})$ is a ``potential energy" term.  Also, we assume $M_{\mu\nu}^{-1}$ is a diagonal fictitious mass matrix which accounts for the possible wide dynamical range in the parameters.  As a first approximation, and without loss of generality, we can assume that the mass matrix is represented by the identity matrix, i.e. $M_{\mu\nu}^{-1}=I_{\mu\nu}$.  The Hamiltonian has a canonical distribution over the phase space of 
\bea
\pi(\vec{x},p^{\mu}) & = & C_{{\mathcal H}}e^{-{\mathcal H}(q^{\mu},p^{\mu})}\\
                              & = & \left(C_E\,e^{-\beta E(q^{\mu})}\right)\left(C_K\,e^{-K(p^{\mu})}\right)\\
                              & = & \pi(q^{\mu}).\pi(p^{\mu}).
\eea
Not only is the above density seperable, but by definition $\pi(p^{\mu})\sim{\mathcal N}(0,1)$, meaning that the momentum components are independent of the state space variables $q^{\mu}$ and each other.  Thus, the marginal distribution for $q^{\mu}$, simply ignoring the momentum variables, gives us a sample set which asymptotically comes from the target distribution.

\subsection{Evolving the Hamiltonian Dynamics.}
To define the dynamics in the phase space, we regard the quantities $(q^{\mu},p^{\mu})$ as being functions of a fictitious time variable $\tau$, which satisfies the dynamical equations
\bea
\frac{dq^{\mu}}{d\tau}=\dot{q}^{\mu} & = & \frac{\partial {\mathcal H}}{\partial p^{\mu}}=p^{\mu} ,\\
\frac{dp^{\mu}}{d\tau}=\dot{p}^{\mu} & = & -\frac{\partial {\mathcal H}}{\partial q^{\mu}}=\frac{\partial \ln {\mathcal L}(q^{\mu})}{\partial q^{\mu}}.
\eea
As with a true physical system, the above equations conserve the total energy of the system (i.e. $d{\mathcal H}/d\tau = 0$).  As the Hamiltonian equations are time reversible (i.e. invariant under the transformation $t\rightarrow-t, p^{\mu}\rightarrow-p^{\mu}$), volume conserving (via Liouville's theorem) and energy preserving, the resultant evolution leaves $\pi(q^{\mu},p^{\mu})$ invariant.
However, in practice, we do not solve the continuous Hamiltonian equations and are forced to solve the Hamiltonian dynamics in a discretized manner.  While there are a number of ways of doing this, caution should be exercised as using the incorrect method destroys both the time reversibility and volume preservation of the method.  One of the most common methods used is a symplectic integrator called the {\em leapfrog integrator} with timestep $\epsilon$.  The algorithm works by first taking a half step in the momenta, followed by a full step in the state parameters, followed by a completing half step in the momenta, i.e. 
\bea
p^{\mu}(\tau+\epsilon/2)&=&p^{\mu}(\tau)+\frac{\epsilon}{2}\left.\frac{\partial \ln {\mathcal L}}{\partial q^{\mu}}\right|_{q^{\mu}(\tau)}\nonumber \\ \nonumber \\
q^{\mu}(\tau+\epsilon)&=&q^{\mu}(\tau)+\epsilon p^{\mu}(\tau+\epsilon/2)\nonumber\\ \nonumber\\
p^{\mu}(\tau+\epsilon)&=&p^{\mu}(\tau+\epsilon/2)+\frac{\epsilon}{2}\left.\frac{\partial \ln {\mathcal L}}{\partial x_i}\right|_{q^{\mu}(\tau +\epsilon)}.
\label{eq:dynamics}
\eea
While the leapfrog algorithm preserves time reversibility and volume conservation, it does not completely conserve the Hamiltonian, introducing an error on the order of ${\mathcal O}(\epsilon^2)$.  In order to make the algorithm ``exact", it was suggested by Duane \emph{et al} to introduce a Metropolis step at the end of each trajectory.  The HMC thus works as follows : 
\begin{enumerate}
\item Generate an initial momentum vector $p^{\mu}_0$ from a Gaussian distribution $\pi(p^{\mu})\sim{\mathcal N}(0,1)$.
\item Starting with an initial point in phase space $(q^{\mu}_0, p^{\mu}_0)$, run the leapfrog algorithm for $l$ steps to reach a new phase space configuration $(q^{\mu}, p^{\mu})$.
\item Accept the new state with probability $\alpha = min\left(exp\left[-\left\{{\mathcal H}'(q^{\mu}, p^{\mu})-{\mathcal H}(q^{\mu}_0, p^{\mu}_0)\right\}\right]\right)$.
\end{enumerate}
As an aside, we should also note that in some fields this algorithm is referred to as a Hybrid Monte Carlo as we are combining a Gibbs sampling of the momenta, followed by a dynamical evolution of the positions, finishing with a Metropolis step.   However, due to the dynamical nature of the problem, we will continue to use the title Hamiltonian Monte Carlo.
\begin{figure}[t]
\begin{center}
\epsfig{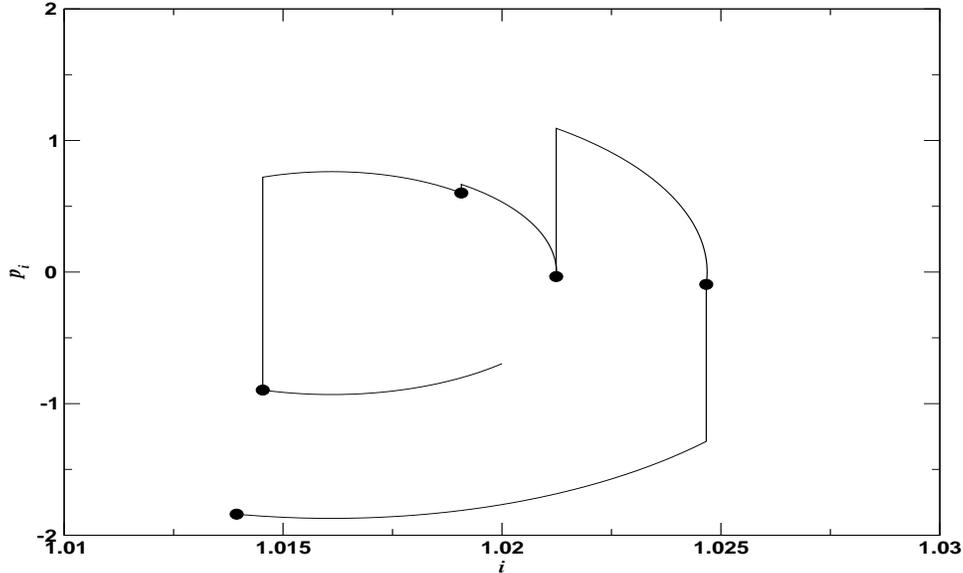}
\end{center}
\caption{Trajectories in inclination phase space $(i, p_i)$.  Each trajectory creates an arc in phase space, while the re-draw of momenta at the end of a trajectory, represented by a solid dot,  corresponds to a vertical jump in the $p_i$ coordinate.  This leads to a new Hamiltonian at the beginning of each trajectory.}
\label{fig:trajectories}
\end{figure}

In Figure~(\ref{fig:trajectories}) we plot several trajectories in inclination phase space $(i, p_i)$, while keeping all other waveform parameters constant.  Each trajectory creates an arc in phase space.  We can see that theoretically, it is possible that if a trajectory is too long, we could end up back at the same starting point in phase space leading to circular orbits.  However, as we can randomise the length of a trajectory, we can always avoid this situation.  In the figure, the re-draw of momentum at the end of a trajectory (represented by a solid dot) corresponds to the vertical jump in the $p_i$ coordinate.  This leads to a new Hamiltonian at the beginning of each trajectory and ensures that the Hamiltonian trajectories fully explore phase space.  We also note that $p^{\mu}$ can be negative allowing trajectories in all directions.

\section{Improving the efficiency of the HMC algorithm.}\label{sec:hmc}
The HMC is dependent on certain internal parameters that need to be tuned to the problem at hand.  As with many such algorithms, there is no
global methodology for how we can best choose these parameters.  For the HMC, the two main parameters to be tuned are the stepsize $\epsilon$ and the number of leapfrog steps $l$, such that the total trajectory length in phase space is $L=l\epsilon$.  In the next few sections we discuss ways
of tuning these parameters and increasing the overall efficiency of the algorithm.

\subsection{Tuning $l$ and $\epsilon$.}
It can be shown that the error in a single leapfrog step is of the order ${\mathcal O}\left(\epsilon^3\right)$.  As the number of leapfrogs for a trajectory is proportional
to $1/\epsilon$, this means that the total error along a trajectory in phase space scales as ${\mathcal O}\left(\epsilon^2\right)$.  If we choose $\epsilon$
to be large, this means that the Hamiltonian is no longer conserved, which results in a low acceptance rate.  However, if $\epsilon$ or $l$ is small, the Hamiltonian remains almost conserved, but we require many calculations of the gradient of the log likelihood, which slows down the algorithm.  Finally, if $l$ is large, we have a reduced exploration, leading to circular trajectories in phase space or a random walk in parameter space.

As a first test, we took $l=100$ and $\epsilon=10^{-3}$.  We demanded that the optimal value of $\epsilon$, keeping $l$ constant would result in
an acceptance rate of $\geq 60\%$.  After each 1000 trajectories we made $\epsilon$ smaller by an order of magnitude in order to achieve the 
acceptance rate criterion.  Our initial value of $\epsilon$ was too big and the chain did not move.  However, once we reached a value of $\epsilon=10^{-6}$, the chain moved with an 
acceptance rate of 87$\%$.  However, when we compared the width of the exploration of the chain to the error ellipse predicted by the FIM, we found that it was far too small.

This spurred us to attack the problem from a different direction.  We realised that the issue with using a single value of $\epsilon$ was the the 
dynamical range of the problem at hand.  If we look at the one sigma error predictions from the FIM, we can see that for the particular source we 
were investigating, $\sigma_{\varphi_c}\sim 1$, while $\sigma_{t_c}\sim 10^{-5}$.  This means that the Hamiltonian is never conserved as the trajectories always take us too far in the narrow coordinate directions.  Because of this we abandoned the idea of using $I_{\mu\nu}$ as the diagonal mass matrix and introduced 
the mass matrix $M_{\mu\nu}^{-1} = s^{\mu}\delta^{\mu\nu}$, where $s^{\mu}$ represents a length scale in the direction of parameter $q^{\mu}$ and $\delta^{\mu\nu}$ is the Kronecker delta.  By re-writing $\epsilon^{\mu}=s^{\mu}\epsilon$ and $\tilde{p}^{\mu} = s^{\mu}p^{\mu}$, we can now re-express Equations~(\ref{eq:dynamics}) as~\cite{Neal}
\bea
\tilde{p}^{\mu}(\tau+\epsilon^{\mu}/2)&=&\tilde{p}^{\mu}(\tau)+\frac{\epsilon^{\mu}}{2}\left.\frac{\partial \ln {\mathcal L}}{\partial q^{\mu}}\right|_{q^{\mu}(\tau)}\nonumber \\ \nonumber \\
q^{\mu}(\tau+\epsilon^{\mu})&=&q^{\mu}(\tau)+\epsilon^{\mu} \tilde{p}^{\mu}(\tau+\epsilon^{\mu}/2)\nonumber\\ \nonumber\\
\tilde{p}^{\mu}(\tau+\epsilon^{\mu})&=&\tilde{p}^{\mu}(\tau+\epsilon^{\mu}/2)+\frac{\epsilon}{2}\left.\frac{\partial \ln {\mathcal L}}{\partial x_i}\right|_{q^{\mu}(\tau +\epsilon^{\mu})}.
\label{eq:moddynamics}
\eea
We should point out here that the successive values of $(q^{\mu}, \tilde{p}^{\mu})$ are no longer Hamiltonian trajectories at constant times, but can still be
used without any loss of generality due to the Metropolis step at the end of the trajectory.  As before, we still chose the initial values of $\tilde{p}^{\mu}$ from $\tilde{p}^{\mu} = {\mathcal N}(0,1)$.  Furthermore,
as with $\epsilon$ and $l$, there is no general manner of choosing the scaling $s^{\mu}$.  From the problem at hand, we identified three possible
ways of doing this : the first is to use the curvature of the log likelihood in the parameter directions (i.e. evaluating elements of the Hessian of the log likelihood), the second is to use the one sigma predictions
from the FIM and the third is to use the eigenvalues from the inverse of the FIM.  Running a 100 trajectory chain with $l=100$ and $\epsilon=5\times10^{-3}$, we obtained acceptance rates
of 69, 65 and 72\% respectively using the three methods.  However, when we examined the exploration, we found that the scaling with the one sigma FIM predictions offered
the widest exploration.

Our next goal was to increase the acceptance rate.  As stated earlier, the acceptance rate is linked to the error in the conservation of the Hamiltonian as we evolve the trajectory.  If the error in the dynamics truly is solely due to the discretisation of the trajectory, i.e. ${\mathcal O}(\epsilon^2)$,  as theory predicts, then simply decreasing the size of $\epsilon^{\mu}$ should increase the acceptance rate.  In our initial runs we approximated the gradient of the log likelihood using a one sided numerical derivative, as it required the least number of waveform generations, i.e.
\begin{equation}
\frac{\partial \ln {\mathcal L}}{\partial q^{\mu}} \approx \frac{\ln {\mathcal L}(q^{\mu} + \delta q^{\mu}) - \ln {\mathcal L}(q^{\mu})}{\delta q^{\mu}} + {\mathcal O}(\delta q^{\mu}),
\label{eq:oneside}
\end{equation}
where we used a value of $\delta q^{\mu} = 10^{-6}$.  To test the above hypothesis, we doubled the number of leapfrogs to $l=200$ and halved the step size to $\epsilon=2.5\times10^{-3}$, thus keeping the trajectory lengths $L$ constant.  Surprisingly we found that the acceptance rate did not change, even though $\epsilon$ was now smaller by a factor of two.  This had to mean that the dominant source of error in the Hamiltonian had to come from the approximation for the gradient of the log likelihood.  By simply changing Equation~(\ref{eq:oneside}) to a central difference approximation of the form
\begin{equation}
\frac{\partial \ln {\mathcal L}}{\partial q^{\mu}} \approx \frac{\ln {\mathcal L}(q^{\mu} + \delta q^{\mu}) - \ln {\mathcal L}(q^{\mu} - \delta q^{\mu})}{2\delta q^{\mu}} + {\mathcal O}((\delta q^{\mu})^2),
\label{eq:twoside}
\end{equation}
we were able to increase the acceptance rate from 65\% to 84\%.  In fact, an investigation of the trajectories revealed run-away divergent terms in the 
kinetic energy as being the culprit for many of the trajectories not being accepted. 

\section{Comparing the MCMC and HMC methods.}\label{sec:mvh}
In order to compare the MCMC and HMC methods, we chose to use the SMBHB system that was studied in Ref.~\cite{B3}.  This system has individual masses of $m_1 = 10^7\,M_{\odot}$ and $m_2 = 10^6\,M_{\odot}$ at a redshift of $z=1$.  This corresponds to chirp and reduced mass values of $M_c = 4.93\times10^6\,M_{\odot}$ and $\mu = 1.82\times10^6\,M_{\odot}$ respectively.  For the angular parameters we use $(\theta, \phi, \iota, \varphi_c, \psi) = (1.325, 2.04, 1.02, 2.954, 0.658)$ radians.  The redshift of the source corresponds to a luminosity distance of $D_{L} = 6.63$ Gpc, and we finally take the time of coalescence and time of observation to be $t_c = 0.49$ and $T_{obs} = 0.5$ years.  The system has a SNR of 390 (due to a different noise curve to the one used in the original article).  The reason for choosing this particular system is that it was shown that the FIM overestimates the error prediction by approximately a factor of two in some parameters.  This is due to the extremely high parameter correlations, which gives us the chance to also test the algorithm in a particularily difficult case.  

\begin{figure}[t]
\begin{center}
\epsfig{file=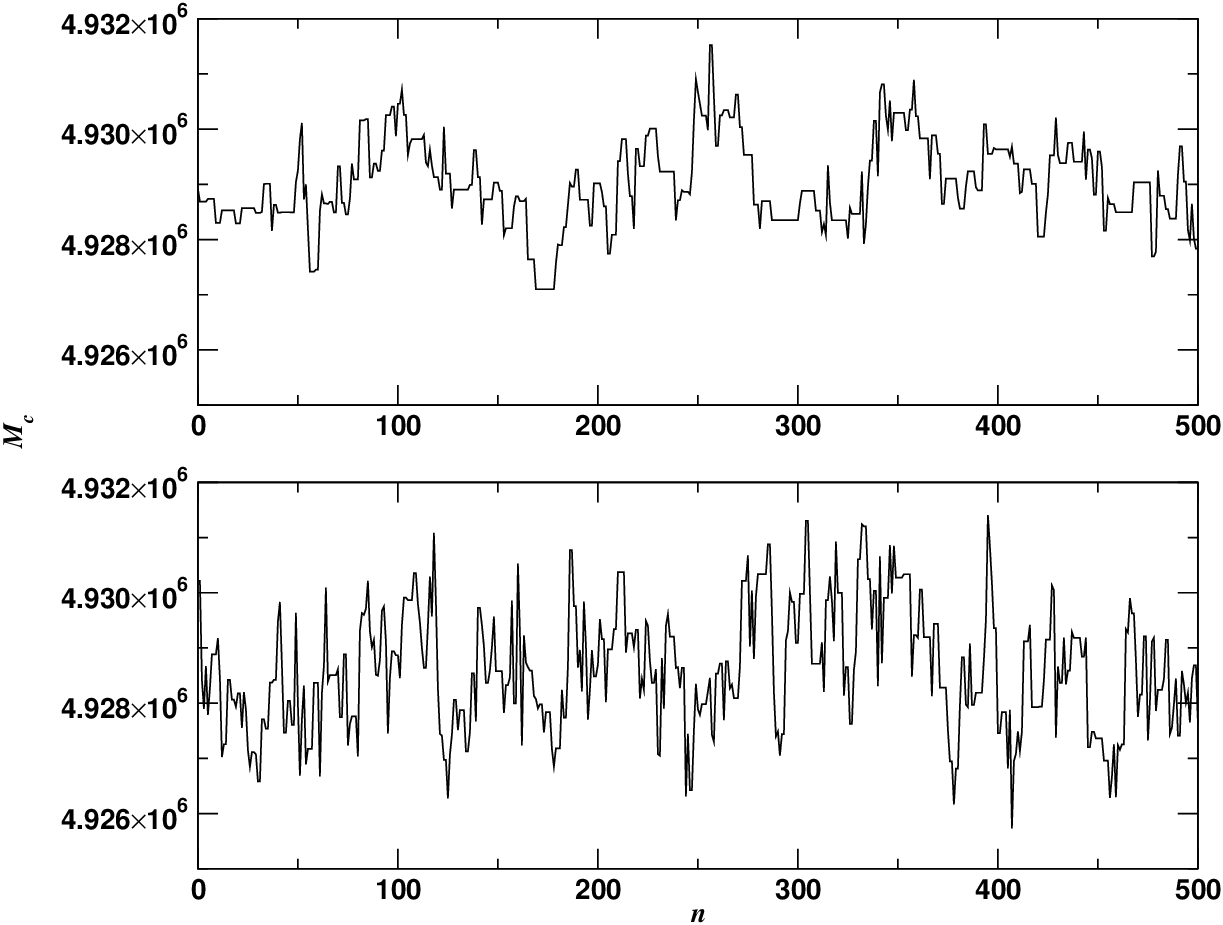, width=5.5in, height=3in}
\caption{We compare a MCMC (top) with a HMC (bottom) over the first 500 iterations of the chains.  We can see a much higher level of exploration with the HMC.  The acceptance rates are 43.8\% and 67.4\% respectively.}
\label{fig:shrtchn}
\vspace{1cm}
\epsfig{file=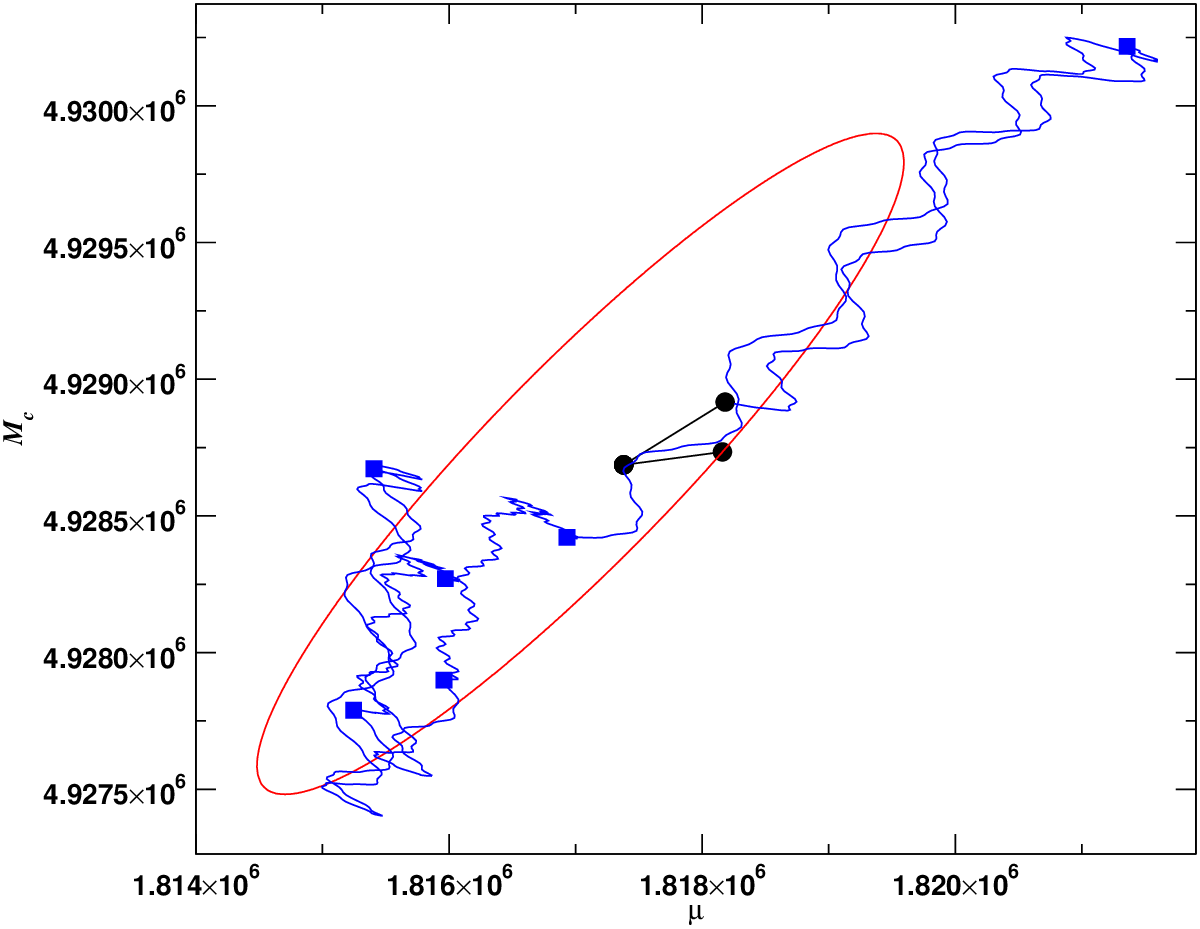, width=5.5in, height=3in}
\end{center}
\caption{A comparison of accepted points from the first six proposals for both a MCMC (circles) and a HMC (squares) algorithm in $(M_c,\mu)$ space.  Only two proposals are accepted for the MCMC, while all six are accepted for the HMC.  In comparison with the 1$\sigma$ error ellipse predicted by the FIM, we can see that the HMC has a already visited both sides of the distribution.}
\label{fig:ellipse}
\end{figure}

To gauge the performance between the MCMC and the HMC, we ran a short 500 iteration chain for both methods.  At the end of the runs the acceptance rates were 43.8\% and 67.4\% respectively.  In Figure~(\ref{fig:shrtchn}) we plot the chirp mass chains for both methods.  We observe a large number of flat spots in the MCMC chain (top).  These represent periods where the chain remained stationary while waiting for a proposal to be accepted We can see that the HMC (bottom cell) has a much larger and faster exploration of the parameter space than the MCMC.  The chains for the other parameters produce similar images.  In Figure~(\ref{fig:ellipse}) we demonstrate this further.  Here we plot the results of the first six iterations of the chains in $(M_c,\mu)$ space.  For the first six iterations of the MCMC, only two out of six proposals were accepted (black circles).  On the other hand, all six proposals were accepted for the HMC (blue squares).  What is most remarkable from this image, is that if we compare with the 1$\sigma$ error ellipse as predicted by the FIM, the exploration of the HMC has already visited both sides of the distribution in just six iterations.  Thus, not only do we have a higher acceptance rate, but we have much wider exploration of the parameter space.   Furthermore, as the MCMC jumps from point to point and has only realistically visited six points in the parameter space. On the other hand, the HMC is following trajectories and visited 1200 points in the parameter space in just six trajectories.  One final thing that we bring attention to, in this image, is the oscillatory nature of the trajectories in the HMC.  This is simply due to the fact that some of the highly correlated parameters have very narrow error ellipsoids in that particular direction.  That means that, again if we picture the inverted likelihood surface as a ``potential well", the algorithm moves in these particular directions by skimming the walls of the potential.  We can thus think of the trajectory as essentially bouncing of the steep walls of the potential as it moves in certain directions.

\subsection{Reducing the Gradient Bottleneck.}
So in terms of exploration, the HMC is clearly superior to a standard MCMC mostly due to the fact that we are using gradient information from the inverse log-likelihood surface in order to dynamically evolve $(q^{\mu}, p^{\mu})$.  However, in practice, the HMC in its basic form is computationally more costly, to the extent that any benefits in convergence are offset by the run-time of the pipeline.  In a standard MCMC, during the phase where we are trying to map the \newpage
\noindent posterior density function, we are essentially sitting on the global maximum in the parameter space.  As we showed earlier, the most efficient way to evolve the MCMC is to use jumps in the eigencoordinates  of the global maximum, whose magnitudes are scaled by the eigenvalues of the FIM.  In general, the MCMC stays within 2-3$\sigma$ of the global 
 solution during the exploration, meaning that the values of the eigenvalues and eigenvectors of the FIM do not change very much.  With this in mind, one could envisage updating the eigenvectors and eigenvalues every few hundred steps just to ensure a good mixing of the chain.  

We see now from Equation~(\ref{eq:FIM}) that calculation of the FIM involves derivatives of the waveforms.  For a $D$ dimensional problem, and working in the easiest case, where we use a forward differencing to calculate the numerical derivative (similar to Equation~(\ref{eq:oneside})), we require $D$ waveform generations (the reason it is not $D+1$ is that the derivative with respect to luminosity distance, $\partial h/\partial \ln D_L = -h$, meaning that we spare ourselves a waveform generation through a scalar multiplication).  If we required a more exact numerical derivative, and used a central difference method similar to Equation~(\ref{eq:twoside}), the number of waveform generations would increase to $2(D-1)+1$ per FIM calculation.  So, for a MCMC of $N$ iterations, using one waveform for the Metropolis-Hastings ratio and updating the eigenvalues and eigenvectors every $n$ steps, the total number of waveform generations per channel is approximately 
\begin{equation}
{\mathcal N} \approx N\left(1+\frac{1}{n}(2(D-1) + 1)\right)
\end{equation}
For the HMC, things are very different.  We can see from Equations~(\ref{eq:dynamics}) that the dynamical evolution requires a calculation of the derivative of the log likelihood at each step of the trajectory.  As we have shown that a central difference numerical derivative for the log likelihood is needed to keep the error in the Hamiltonian minimised, we require $2(D-1) + 1$ waveforms per leapfrog step (again due to the fact that we can treat derivatives with respect to the luminosity distance as a scalar multiplication), which give us $l\times (2(D-1)+1)$ waveforms per trajectory.  If we have $N$ trajectories (or iterations in the MCMC), the total number of waveforms needed per channel is 
\begin{equation}
{\mathcal N} \approx N\left(1 + l(2(D-1) + 1)\right).
\end{equation}
As $Nl\gg N/n$, we can see that the main bottleneck in the code is the calculation of the gradient of the log-likelihood.  To give a concrete example : if we assume a MCMC of $N=10^6$ iterations, where we update the FIM every $n=10$ 
\noindent iterations, and assume non-spinning waveforms where $D=9$, the total number of waveform generations is $5.4\times10^6$, assuming a two channel detector.    Now if we assume that we have the same length HMC algorithm, with $l=200$ leapfrogs per trajectory, the total number of waveform generations is $6.8\times10^9$.  A further problem is that if we try to cut down the number of times the gradient is calculated by reducing $L$, or only updating the gradient every $m$ steps in a single trajectory, we get a runaway kinetic energy term in the Hamiltonian and the phase space transitions are never accepted.  This demonstrates quite clearly that while we acquire greater exploration with the HMC, in its standard form, the computational cost outways any other benefits. 

To try and circumvent this problem, we investigated the possibility of using an analytic function to model the gradient of the log likelihood.  To begin with we tried fitting both a multivariate Gaussian and a Skew-Normal distribution.  However, in both cases we could not fit the data sufficiently well to move the chains.  We observed that because of the high correlations between parameters, and the fact the in certain directions the distributions are highly skewed, the analytic functions were too constrained to provide a good fit.  

However, we can see from Equations~(\ref{eq:moddynamics}) that the gradients of the log likelihood are only in the coordinate directions themselves.  This gave us the idea of fitting each of the nine gradients independently using a cubic function of the form
\begin{equation}
f(q^{\mu}) = a_0 + \sum_{i=1}^D a_i q^i + \sum_{j=1}^D \sum_{k=j}^D a_r q^j q^k + \sum_{l=1}^D \sum_{v=l}^D \sum_{w=v}^D a_s q^l q^v q^w,
\end{equation}
where $D$ is the number of waveform parameters.  In the above equation, we have written $a_r$ to represent the independent coefficients of the matrix $a_{jk}$.  This is a $D\times D$ symmetric matrix with 45 independent coefficients for non-spinning binaries.  Similarily, we use $a_s$ to represent the $D\times D\times D$ symmetric matix $a_{lvw}$.  This matrix has 165 idenpendent components.  In total, this means that for each
of the nine gradients we need to fit 220 coeficients.  As this function is linear in the coefficients $\vec{a}$ we can use a least squares algorithm to solve for the coefficients.  So, writing the least squares function in the form
\beq
S(\vec{a}) = \sum_{i=1}^N \left (\sum_{j=0}^M a_j \phi_j(q^{\mu}_i) - y_i\right)^2,
\eeq
where $N$ denotes the number of data points, there are $M+1$ fitting coefficients and the functions $y_i$ and $\phi_{j}({q^{\mu}})$ are defined by
\beq
y_i\equiv  \left. \frac{\partial \ln \mathcal{L}}{\partial q^{\mu}}\right\vert_i, \,\,\,\,\,\,\,\,\,\phi_{j}({q^{\mu}}) \equiv \frac{\partial f(q^{\mu}) }{\partial a_j}.
\eeq
It is more convenient to write this in matrix form as
\beq
S(\vec{a}) = (\bf{J}\vec{a}-\vec{y})^T(\bf{J}\vec{a}-\vec{y}),
\label{eq:matrix_S}
\eeq
\newpage
\begin{figure}[t]
\begin{center}
\epsfig{file=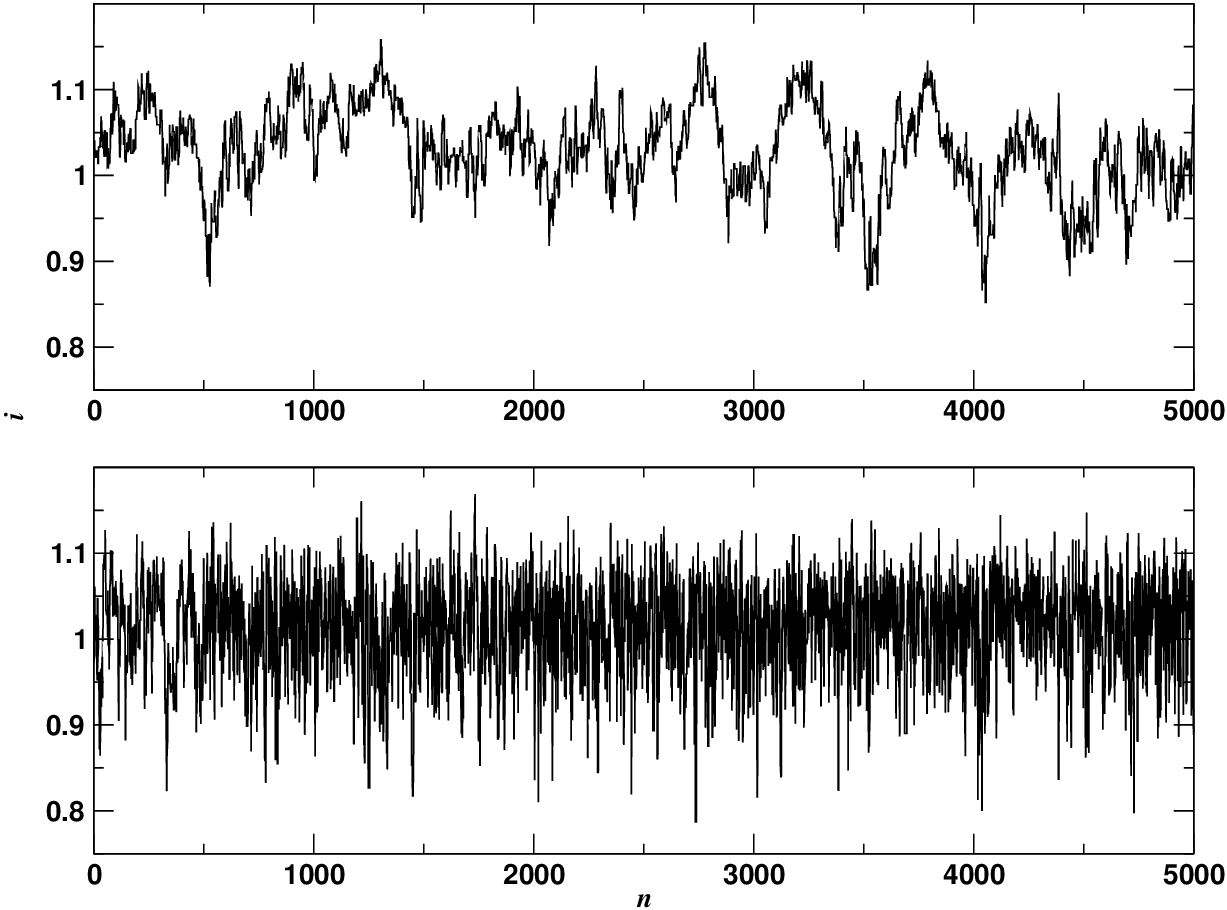, width=5.5in, height=2.8in}
\caption{We compare a MCMC (top) with a HMC (bottom) over the first 5000 iterations of the chains for the inclination.  We can see a much higher level of exploration with the HMC.  The acceptance rates are 46\% and 73\% respectively.}
\label{fig:i5000}
\vspace{1cm}
\epsfig{file=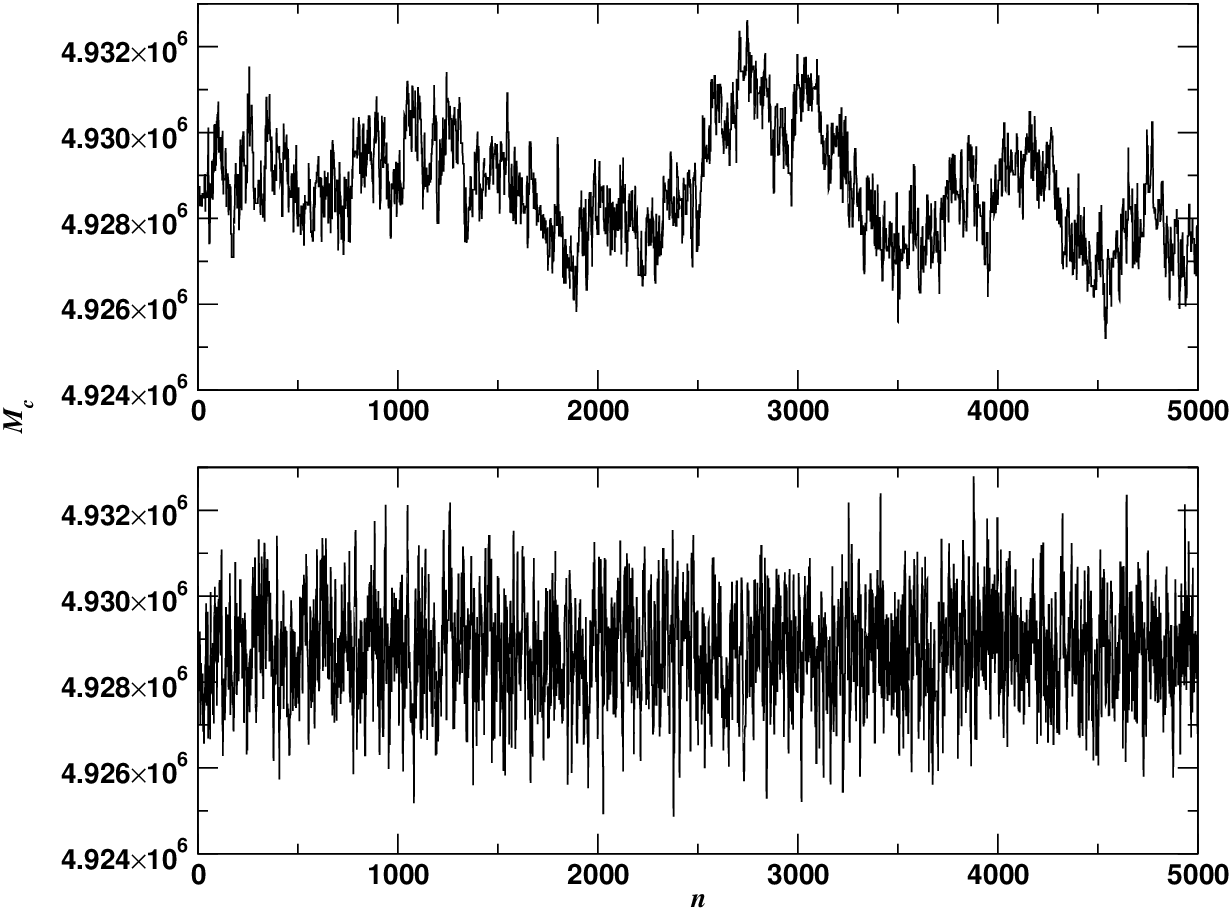, width=6in, height=2.8in}
\end{center}
\caption{The same representation for the chirp mass.}
\label{fig:mc5000}
\end{figure}
\noindent where
\beq
\bf{J}=\left( \begin{array}{c c c}
\phi_0(q^{\mu}_1) & \ldots & \phi_{M}(q^{\mu}_1)\\
\vdots & \ddots & \vdots\\
 \phi_0(q^{\mu}_N) & \ldots & \phi_{M}(q^{\mu}_N)\end{array} \right),
 \,\,\,\, \vec{a} = \left( \begin{array}{c}
 a_0\\
 \vdots\\
 a_M \end{array} \right),
  \,\,\,\, \vec{y} = \left( \begin{array}{c}
 y_1\\
 \vdots\\
 y_N \end{array}
 \right).
\eeq
Here $\bold{J}$ is the Jacobian matrix. To find the fitting parameters, we solve the following set of equations
\beq
\frac{\partial S(\vec{a})}{\partial a_k}=0,
\eeq
which, combined with Equation (\ref{eq:matrix_S}), gives us
\beq
2\bold{J^TJ}\vec{a}-2\bold{J^T}\vec{y}=\vec{0}.
\eeq
Now, solving for $\vec{a}$ gives the solution
\beq
\vec{a}_{fit}=(\bold{J^TJ})^{-1}\bold{J^T}\vec{y}.
\eeq
To fit the data, we only use the trajectory points from accepted trajectories.  The reason for this is that sometimes a trajectory visits a point in parameter
space where the log likelihood gradient is numerically unstable.  These trajectories are never accepted as they do not conserve the Hamiltonian and
thus are rejected.  We tried fitting the gradients after differing numbers of trajectories and found that after 500 trajectories the improvement in the 
$a_i$ parameters was less than the tolerance we defined for the fit.  The least squares fit took approximately 50 minutes to fit all nine log likelihood
gradients.

\section{Applying the Hamiltonian Monte Carlo to supermassive black hole binaries.}\label{sec:results}
Using the binary system defined in the previous section, we compared both the MCMC and HMC algorithms using a $10^6$ iteration chain.  For the 
MCMC we updated the FIM every 20 iterations to ensure a constant mixing of the chain.  For the HMC we used the following setup : 
\begin{enumerate}
\item Set $\epsilon=2.5\times10^{-3}$ and $l=200$,
\item	Start the HMC using two-sided numerical derivates for the log likelihood gradients, and use a scaling of 
$s^{\mu}=\sigma^{\mu}_{FIM}=\Gamma_{\mu\mu}^{-1/2}$ in the dynamical equations,
\item After 500 trajectories, use the data points from the accepted trajectories to analytically fit the nine log likelihood gradients,
\item Reset the scaling to $s^{\mu} = \sigma^{\mu}_{chain}$, and double the number of leapfrogs per trajectory to $l=400$.
\end{enumerate}
At the end of the two runs, we achieved acceptance rates of 46\% and 73\% respectively for the MCMC and HMC algorithms.  As a point of 
comparison, we also ran a short HMC where we used full numerical derivatives for the log likelihood gradients.  This chain had an acceptance
rate of 65\% at the point of comparison, but was taking too long to complete the full $10^6$ iterations, so the run was stopped after 37,000 iterations.

In Figures~(\ref{fig:i5000}) and (\ref{fig:mc5000}) we plot the first 5000 chain points for the parameters $\iota$ (top) and $M_c$ (bottom).  In each figure, the chain for the MCMC is represented in the top cells, while the HMC chains are presented in the bottom cells.  For both parameters we can
see that there is a high level of random walk behaviour in the MCMC chains.  This is particularly evident with the chirp mass where we see that the 
overall oscillatory evolution of the chain is more prominant than the small scale fluctuations of the chain.  This suggests that the chain is mixing very
slowly in this parameter direction.  However, if we refer to the lower cells of each figure, we can see that the random walk behaviour has disappeared
in the case of the HMC where we have excellent mixing of the chain, and more importantly, that the chain is moving across the entire solution quite rapidly.

\begin{figure}[t]
\begin{center}
\epsfig{file=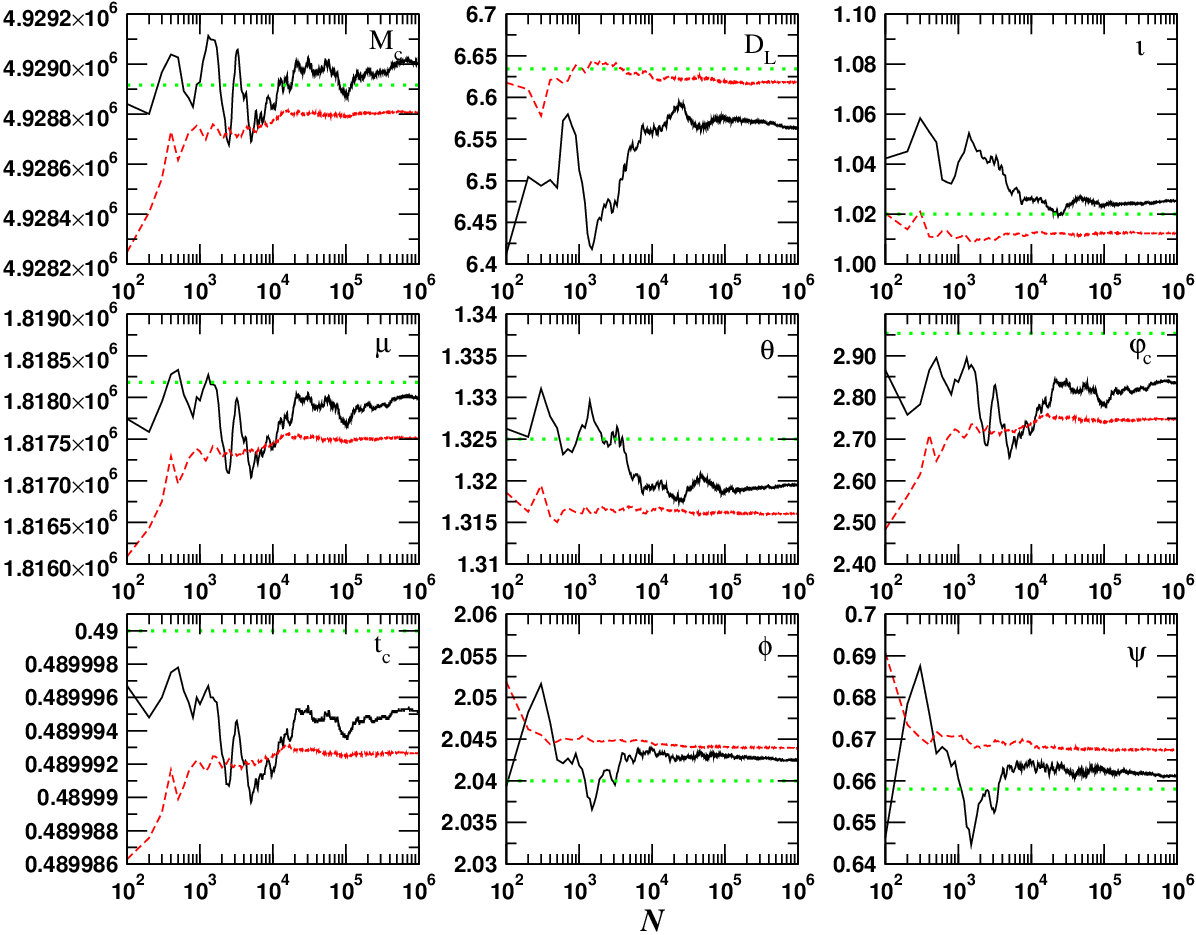, width=5.7in, height=3in}
\caption{We compare the instantaneous chain means for a MCMC (black - solid) with a HMC (red - dashed).  The injected values are represented by the flat green dashed lines.    We can see that the MCMC means take a lot longer to stablise, while the HMC means flatten very quickly demonstrated rapid convergence.}
\label{fig:instmeans}
\vspace{1cm}
\epsfig{file=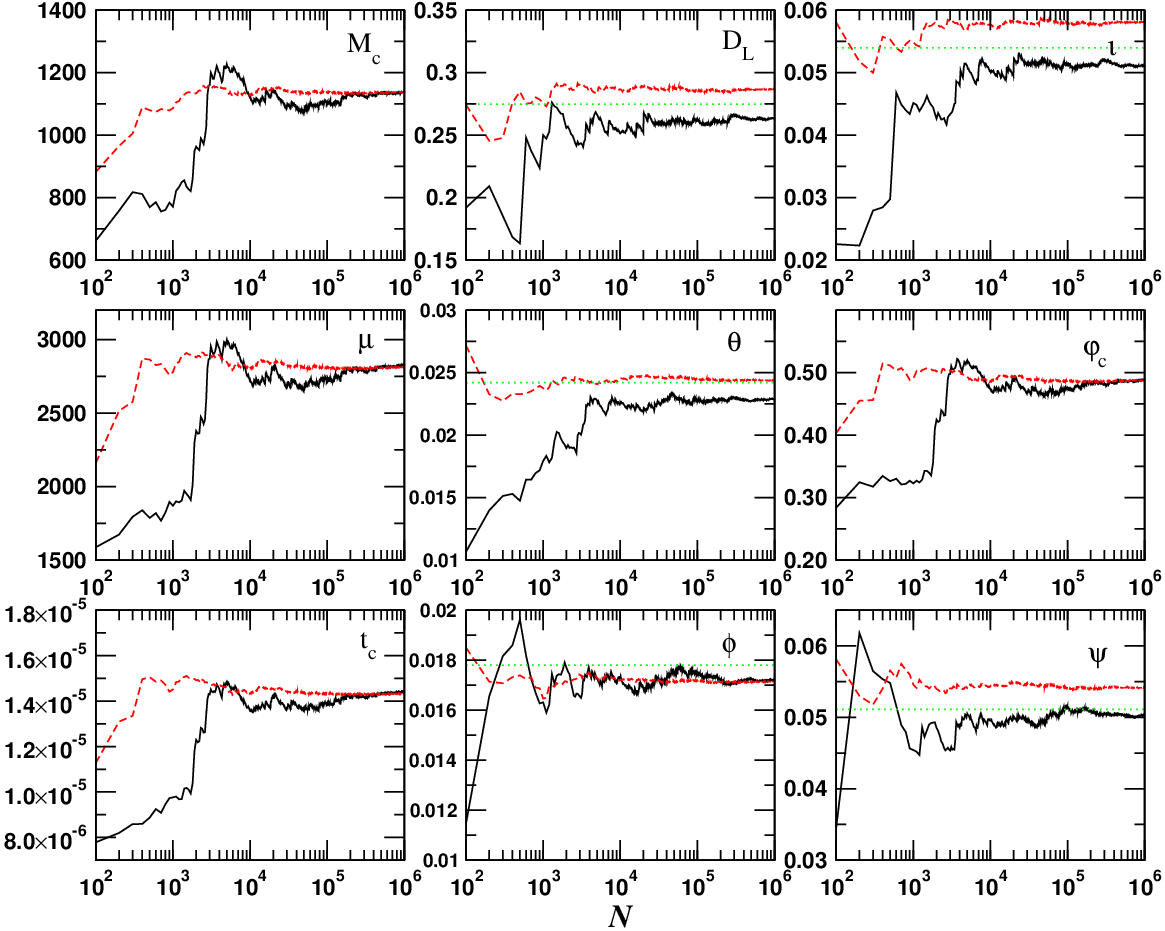, width=5.5in, height=3in}
\end{center}
\caption{We compare the instantaneous standard deviations for a MCMC (black - solid) with a HMC (red - dashed).  The FIM predictions are represented by the flat green dashed lines.    Once again,  the HMC means flatten very quickly demonstrated rapid convergence.}
\label{fig:inststdevs}
\end{figure}

One way to test the convergence of a Markov chain is to visually inspect the instantaneous means and standard deviations from the chain (i.e. at each iteration of the chain we re-calculate the means and standard deviations given the totality of information), and investigate how quickly major oscillations dissipate and the curves flatten.  In Figure~(\ref{fig:instmeans}) we plot the instantaneous means for the two algorithms.  The MCMC chains are represented by the solid
(black) curves, while the HMC is represented by the dashed (red) curves.  The flat (dashed green) lines represent the injected values.  The first thing that we would
like to emphasise here, is that given the finite length of the chains, we do not expect both algorithms to converge to exactly the same value.  If the chains were
run for an infinitely long time, we would expect perfect agreement on the order of numerical error.  That said, it is also important to note that the chains for both
algorithms return final means that are within $0.06 \leq \sigma \leq 0.5$ of the injected values, displaying remarkable precision.  The main aspect that we would
like the reader to take away from this figure is how quickly major oscillations dissipate in the case of the HMC chains and how fast they flatten.  This displays a 
highly accelerated convergence.  Once again, we point out that it has been shown that a HMC which is $D$ times shorter than a standard MCMC chain will yield
the same performance~\cite{Hajian}, so this accelerated convergence is to be expected.

If we now look at the instantaneous standard deviations in Figure~(\ref{fig:inststdevs}), the representations of the curves are identical to the previous image.  The
sole difference is that the flat lines now represent the $1\sigma$ prediction from the FIM.  In some cases this line is missing as the FIM overestimates the error
in these parameters.  We observe a similar behaviour in terms of convergence to the means.  The HMC curves once again flatten much quicker than the MCMC
curves.

\begin{figure}[t]
\begin{center}
\epsfig{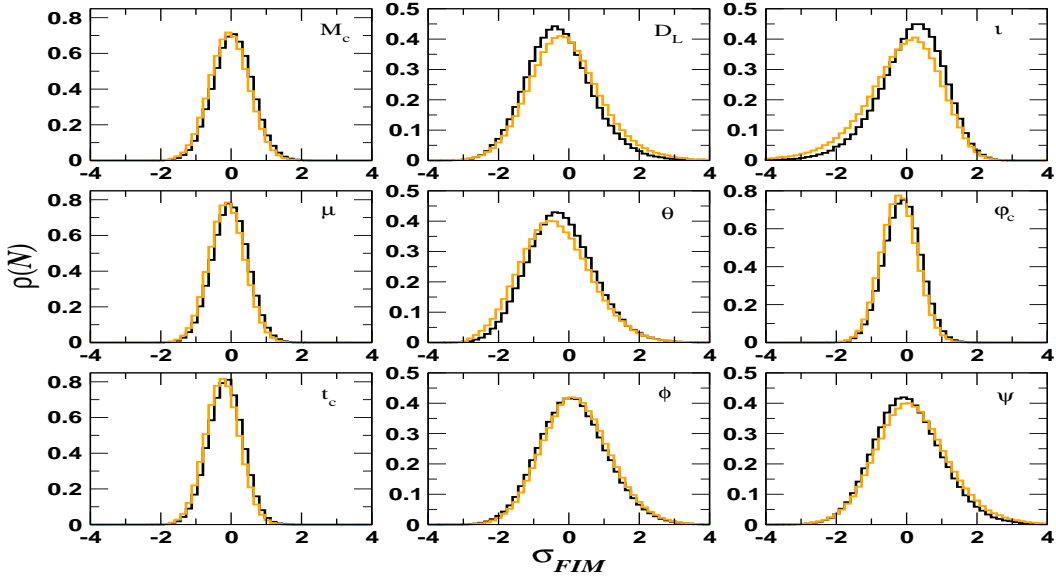}
\end{center}
\caption{A comparison of histograms from the MCMC (black) and HMC (orange) chains.}
\label{fig:hists}
\end{figure}

In Figure~(\ref{fig:hists}) we plot histograms from the chains based on the two algorithms.  The histograms from the MCMC algorithm are represented by the 
dark (black) curves, while the HMC histograms are represented by the light (orange) 
\newpage
\noindent curves.  We can see that both chains capture the skewed distributions for 
parameters like $\iota$ and $D_L$, and the narrow distributions for $M_c, \mu, t_c$ and $\varphi_c$.  To further test the hypothesis of a $D$ times acceleration
in convergence, we ran the MCMC for $10^7$ iterations.  We then recalculated the histograms at $10^5$ steps from $8\times 10^6$ to $10^7$ iterations.  We 
found that for a $8.7\times 10^6$ iteration MCMC chain, the MCMC histograms are virtually indistinguishable from the HMC histograms, supporting the factor
of $D$ in efficiency.  Finally, In Table~(\ref{tab:res}) we provide the means, medians and standard deviations from both sets of chains, as well as the injected parameter values and the
$1\sigma$ predictions from the FIM.

\begin{table}[h]
\centering
\begin{tabular}{c c c c c c}
\hline
					& true		& mean		& mean	& median	& median\\
					& values		& MCMC		& HMC	& MCMC	& HMC	\\
\hline
$M_c/10^6M_\odot$		& 4.9289		& 4.9290			& 4.9288		& 4.92899			& 4.92880\\
$\mu/10^6M_\odot$		& 1.818182	& 1.817985		& 1.817512	& 1.817987		& 1.817507\\
$t_c/yrs$				& 0.49		& 0.4899951		& 0.4899927	& 0.4899953		& 0.4899927\\
$\theta$				& 1.325		& 1.3195			& 1.3160		& 1.3185			& 1.3149\\
$\varphi$				& 2.04		& 2.04253			& 2.04396		& 2.0419			& 2.0433\\
$D_L/Gpc$			& 6.63433		& 6.56377			& 6.61829		& 6.55007			& 6.59565	\\
$i$					& 1.02		& 1.0253			& 1.0123		& 1.0301			& 1.0187\\
$\phi_c$				& 2.954		& 2.8337			& 2.7477		& 2.8376			& 2.7487\\
$\Psi$				& 0.658		& 0.6613			& 0.6674		& 0.6588			& 0.6644\\				
\hline \hline
			& $\sigma_{MCMC}$			& $\sigma_{HMC}$			& $\sigma_{FIM}$\\
\hline
$M_c$		& 1139.71				& 1137.00				& 2047.91\\
$\mu$		& 2831.65				& 2810.58				& 5582.54\\
$t_c/yrs$		& $1.443\times10^{-5}$	& $1.431\times10^{-05}$	& $2.956\times10^{-5}$\\
$\theta$		& 0.0229				& 0.0244				& 0.0242\\
$\varphi$		& 0.0172				& 0.0171				& 0.0178\\
$D_L/Gpc$	& 0.2633				& 0.2866				& 0.2748\\
$i$			& 0.0521				& 0.0581				& 0.0539\\
$\phi_c$		& 0.4893				& 0.4873				& 0.9369\\
$\Psi$		& 0.0503				& 0.0541				& 0.0511\\				
\hline
\end{tabular}
\caption{Means, medians and standard deviations of the MCMC and the HMC after an one million point chain. We also give the standard deviations predicted by the FIM.}
\label{tab:res}
\end{table}

The final aspect that we would like to compare is the run-time per algorithm.  For the MCMC, we ran a $10^6$ iteration chain, updating the FIM every 20 steps.
Using a MacBook Pro with an 2.66 GHz i7 processor and 8GB of memory, the MCMC takes $\sim$15 hours to run with an acceptance rate of 46\%.  For the 
HMC, we ran $10^6$ trajectories.  For the first 500 trajectories, we used $l=200$ leapfrog steps per trajectory, taking 2 hours and 45 minutes.  The pipeline then
took a further 50 minutes to fit the nine log likelihood gradients.  Once we had an analytic fit for the gradients, we doubled the number of leapfrog steps
to $l=400$, keeping $\epsilon$ constant.  The code then took another 3.5 hours to complete the
$10^6$ trajectories with an acceptance rate of 73\%.  This gives a total run-time for the HMC of $\sim$7.5 hours.

While the HMC still requires a factor of $\sim$1.5 more waveform generations, the MCMC takes almost twice as long to run with $D$ times less
efficiency.  This extra run-time comes mostly from the calculation of the proposal distribution.  Every 20 steps, and assuming a two channel detector,  we generate 34 waveforms for the FIM
calculation, which then requires 34 numerical Fourier Transforms.  As the FIM is a symmetric matrix, this then involves the calculation of 90 inner products, an inversion, and the calculation of the eigenvalues and
eigenvectors.  With the HMC, once we have the analytic fits to the gradients, the run-time cost of a trajectory plus the waveform generation at the end of the 
trajectory is almost equivalent to the run-time cost of the waveform generation, leading to a much faster algorithm.

\section{Conclusion}
In this article we present an implementation of a Hamiltonian Markov Chain algorithm for the parameter estimation of supermassive black hole binaries.  While
the Markov Chain Monte Carlo method has been shown to be highly effective for GW sources, we know that the types of algorithms already implemented belong
to a family of random walk algorithms.  These algorithms need to be carefully tailored to the problem at hand, and due to their random nature, can take a long
time to converge.  They also require long chains in order to reduce the autocorrelation between samples.  The Hamiltonian Markov Chain tries to reduce the random walk
behaviour by introducing a set of state space variables and canonical momenta, and then evolving trajectories in phase space by solving Hamilton's equations.

We have demonstrated that instead of using a single step size in the Hamiltonian equations, by using different step sizes in the different coordinate directions, we improve the acceptance rate and take care of the large condition number of the problem.  While this improved the acceptance rate of the chain, we showed that
the main source of improvement in the acceptance rate was using a central difference equation for the numerical derivatives of the log likelihood.  While now 
more accurate, the execution of the algorithm was slower due to the extra number of waveforms that needed to be generated.  However, we were able to 
circumvent this problem by using the data points visited by accepted trajectories during the first 500 iterations of the algorithm to fit an analytic cubic function that
represents the gradient of the log likelihood in a coordinate direction.  This fit was then accurate enough to use in evolving the trajectories, dramatically reducing 
the number of waveform generations needed.  This resulted in an algorithm that was executed almost twice as fast, and almost 9 times more efficiently than a
 Metropolis-Hastings MCMC of the same length.


\section*{References}
\begin{thebibliography}{99}
\bibitem{eLISA} \emph{The Gravitational Universe}, The eLISA Constortium, Whitepaper submitted to ESA for the L2/L3 Cosmic Vision call. arXiv:1305.5720
\bibitem{B1} \njc and \ekp, \cqg {\bf 24} 5729 (2007)
\bibitem{B2} \njc and \ekp, \prd {\bf 75} 021301 (2007)
\bibitem{B3} \njc and \ekp, \cqg {\bf 23} S761 (2006)
\bibitem{B4} \jrg and \ekp, \cqg {\bf 26} 225004 (2009)
\bibitem{B5} Babak~S, \jrg and \ekp, \cqg {\bf 26} 135004 (2009)
\bibitem{B6} \njc, \cqg {\bf 28} 094016 (2011)
\bibitem{B7} \ekp and Sesana~A, arXiv:1005.5296 (2010)
\bibitem{B8} Feroz~F, \jrg, Hobson~M~A and \ekp, \cqg {\bf 26} 215003 (2009)
\bibitem{B9} Cohen~M~I, Cutler~C and Vallesneri~M, \cqg {\bf 27} 185012 (2010)
\bibitem{B10} \njc and Littenberg~T~B, \prd {\bf 76} 083006 (2007)
\bibitem{B11} Littenber~T~B and \njc, \prd {\bf 80} 064032 (2009)
\bibitem{B12} Stroeer~A and Veitch~J, \prd {\bf 80} 064032 (2009)
\bibitem{B13} Feroz~F, \jrg, Graff~P, Hobson~M~A and Lasenby~A, \cqg {\bf 27} 075010 (2010)
\bibitem{B14} Gossan~S, Veitch~J and Sathyaprakash~B~S, arXiv:1111.5819 (2011)
\bibitem{B15} Trias~M, Veitch~J and Vecchio~A, \cqg {\bf 26} 204024 (2009)
\bibitem{B16} Stroeer~A \emph{et al}, \cqg {\bf 24} S541 (2007)
\bibitem{B17} Amaro-Seoane~P, Eichhorn~C, \ekp and Spurzem~R, MNRAS {\bf 401} 2268 (2010)
\bibitem{B18} \njc, Sampson~L, Yunes~N and Pretorius~F, \prd {\bf 84} 062003 (2011)
\bibitem{B19} Crowder~J and \njc, \prd {\bf 75} 043008 (2007)
\bibitem{B20} Petiteau~A, Shang~Y, Babak~S and Feroz~F, \prd {\bf 81} 104016 (2010)
\bibitem{B21} Brown~D~A, Crowder~J, Cutler~C and Mandel~I, \cqg S595 (2007)
\bibitem{B22} Umst\"ater~R, Christensen~N, Hendry~M \emph{et al}, \prd {\bf 72} 022001 (2005)
\bibitem{B23} Ali~A, Christensen~N, Meyer~R and Rover~C, \cqg {\bf 29} 145014 (2012)
\bibitem{gibbs} Geman~S and Geman~D, IEEE Transactions on Pattern Analysis and Machine Intelligence {\bf 6} 721 (1984)
\bibitem{MH1} Metropolis~N, Rosenbluth~A~W, Rosenbluth~M~N, Teller~A~H and Teller~E, Journal of Chemical Physics {\bf 21} 1087 (1953)
\bibitem{MH2} Hastings~W~K, Biometrika {\bf 57} 97 (1970)
\bibitem{HMC} Duan~S, Kennedy~A~D, Pendleton~B and Roweth~D, Phys. Lett. B {\bf 195} 216 (1987)
\bibitem{Hajian} Hajian~A, \prd {\bf 75} 083525 (2007)
\bibitem{lisa} P.~Bender et al., \emph{LISA pre-phase A report} (1998)
\bibitem{Verdi} Verdi~L \emph{et al}, MNRAS {\bf 349} 181 (2004)
\bibitem{CR} \njc and Rubbo~L, \prd {\bf 67} 022001 (2003)
\bibitem{RPC} Rubbo~L, \njc and Poujade~O, \prd {\bf 69} 082003 (2004)
\bibitem{cc} Cutler~C, Phys. Rev. D {\bf 57}, 7089 (1998)
\bibitem{nyp} Nelemans~G, Yungleson~L~R and Portegies-Zwart~S~F, MNRAS {\bf 349} 181 (2004)
\bibitem{trc} Timpano~S, Rubbo~L and \njc, \prd {\bf 73}, 122001 (2006)
\bibitem{Neal} Neal~R~M,  \emph{MCMC using Hamiltonian dynamics}, in \emph{The Handbook of Markov Chain Monte Carlo}, Brooks~S, Gelman~A, Jones~G~L, and Meng~X~L (editors), Chapman \& Hall, CRC Press,  113 (2010)
\end{thebibliography}
\end{document}